\documentstyle[12pt,epsfig]{ioplppt}          % use this for preprint style
\def\beq{\begin{equation}}
\def\eeq{\end{equation}} 
\def\beqn{\begin{eqnarray}} 
\def\eeqn{\end{eqnarray}} 
\def\as{\alpha_s} 
\def\asmz{\alpha_s(M_Z^2)} 
 
\def\GeV{{\rm GeV}}

\def\asQ{\alpha_s(Q^2)}

\def\c2w{\cos^2{\theta_W}} 
\def\lapproxeq{\lower .7ex\hbox{$\;\stackrel{\textstyle <}{\sim}\;$}} 
\def\gapproxeq{\lower .7ex\hbox{$\;\stackrel{\textstyle >}{\sim}\;$}}

\def\ubar{\bar u} 
\def\dbar{\bar d} 
\def\sbar{\bar s}
\def\cbar{\bar c}

\def\msb{{\overline{\rm MS}}} 
\def\cO{{\cal O}}

\def\asup{{\alpha_s\uparrow\uparrow}}
\def\asdown{{\alpha_s\downarrow\downarrow}}

\def\np#1#2#3{19#3 {\em Nucl.\ Phys.}~B {\bf#1} #2}
\def\pl#1#2#3{19#3 {\em Phys.\ Lett.}~{\bf#1B} #2}

\def\sjnp#1#2#3{19#3 {\em Sov.\ J.\ Nucl.\ Phys.}~{\bf#1} #2}
\def\spj#1#2#3{19#3 {\em Sov.\ Phys.\ JETP}~{\bf#1} #2}

\def\epj#1#2#3{19#3 {\em Eur.\ Phys.\ J.}~C {\bf#1} #2}
\begin{document} 
\begin{flushright}
DTP/99/08 \\
January 1999
\end{flushright}
\title{Theoretical aspects of high--$Q^2$ deep inelastic
scattering\footnote{Plenary talk presented at the 
3rd UK Phenomenology Workshop on HERA Physics, Durham, September 
1998, to be published in the Proceedings.}}[High--$Q^2$ DIS theory]
 
\author{W J Stirling}
 
\address{Departments of Physics and Mathematical Sciences, University of
Durham, South Road, Durham DH1~3LE, United Kingdom}

\begin{abstract} 
We present an overview of the theory of high--$Q^2$ deep inelastic
scattering. We focus in particular on the theoretical uncertainties
in the predictions for neutral and charged current cross sections
obtained by extrapolating from lower $Q^2$.
\end{abstract}
 
% 
%  Uncomment out if preprint format required 
% 
%\pacs{00.00, 20.00, 42.10} 
%\maketitle
 
\section{Introduction}

The measurement of deep inelastic scattering cross sections
$d^2\sigma/dxdQ^2$ at high $Q^2$ provides an incisive test of the
Standard Model. Interesting results have already been obtained by the H1
and ZEUS collaborations at HERA --- a summary of these can be found in
the accompanying experimental review by Mehta \cite{Andy}. In this talk
we will concentrate on some theoretical issues; in particular, how well
can we predict cross sections for neutral (NC) and charged current (CC)
$e^-p$ and $e^+ p$ scattering cross sections at high $Q^2$, and what can
we hope to learn from present and future measurements at HERA? Many of
the issues presented here were subsequently discussed in detail by the
high--$Q^2$ Working Group at the Workshop \cite{WGi},
and as a result some of the issues raised here were clarified. The
importance of the high--$Q^2$ region as a probe of standard and new
physics was repeatedly emphasised in the discussions.

We begin by recalling the form of the DIS cross sections in the Standard
Model:
\beq
{d^2\sigma_{NC,CC}(e^\pm p) \over d x d Q^2}\  = \
\left\{\begin{array}{c} \mbox{standard LO} \\ \mbox{expressions}
\end{array} \right\}\ + \ \delta_{\rm QCD} \ + \ \delta_{\rm EW} \; .
\label{eq:basic}
\eeq
The NC and CC cross sections are obtained from the $ep \to eX$
($\gamma^*,Z^*$ exchange) and $ep \to \nu X$ ($W^*$ exchange) processes
respectively.   In  (\ref{eq:basic}) the first term on the right-hand side
represents the standard leading-order `parton-model'
 expressions for the deep inelastic structure functions ($F_{2,L,3}$).
Note that at the
high $Q^2 > \cO(10^4\; \GeV^2)$ values measured at HERA,
the  $Z^0$--exchange contribution to the NC
structure functions, see below, cannot be neglected.
The second term represents perturbative QCD next-to-leading
order (NLO) corrections, expressions for which can be found in the
literature (see for example Ref.~\cite{DICK}). Away from $x=0$ and $x=1$
these do not have any particularly dramatic effect on the leading-order cross
sections. Since they  are generally automatically included in the
various computer codes used to fit data and
make predictions, they will not be discussed further here.\footnote{Of
course going beyond leading order necessitates a choice of
renormalisation and factorisation scheme. All quantities referred to in
this talk correspond to the $\msb$ scheme.}

The third term on the right-hand side of (\ref{eq:basic}) represents
electroweak radiative corrections. These are known to at least
$\cO(\alpha)$ and
include QED corrections from photon emission off the incoming and
outgoing quarks and leptons, and also genuine electroweak 
corrections from propagator, vertex and box contributions associated with
the electroweak gauge boson exchanges. The latter allow a precise theoretical
definition of the various electroweak parameters (e.g. vector and axial
couplings, gauge boson masses etc.) that appear in the leading-order
expressions. For a comprehensive review, see the contribution by
Spiesberger in \cite{WGi}.

Three essentially separate
types of information can therefore be obtained from high-precision
 measurements of the cross sections (\ref{eq:basic}) at HERA :
\begin{itemize}
\item parton distributions $f_i(x,Q^2)$  (and $\asQ$) at high $x$ and
$Q^2$,
\item electroweak parameters, in particular $M_W$ and 
$G_\mu$, from the space-like $W$ exchange in the CC cross section,
\item limits on, or measurements of, new physics effects (quark
substructure, leptoquark production, etc.).
\end{itemize}
\begin{figure}[htb]
\begin{center}
\mbox{\epsfig{figure=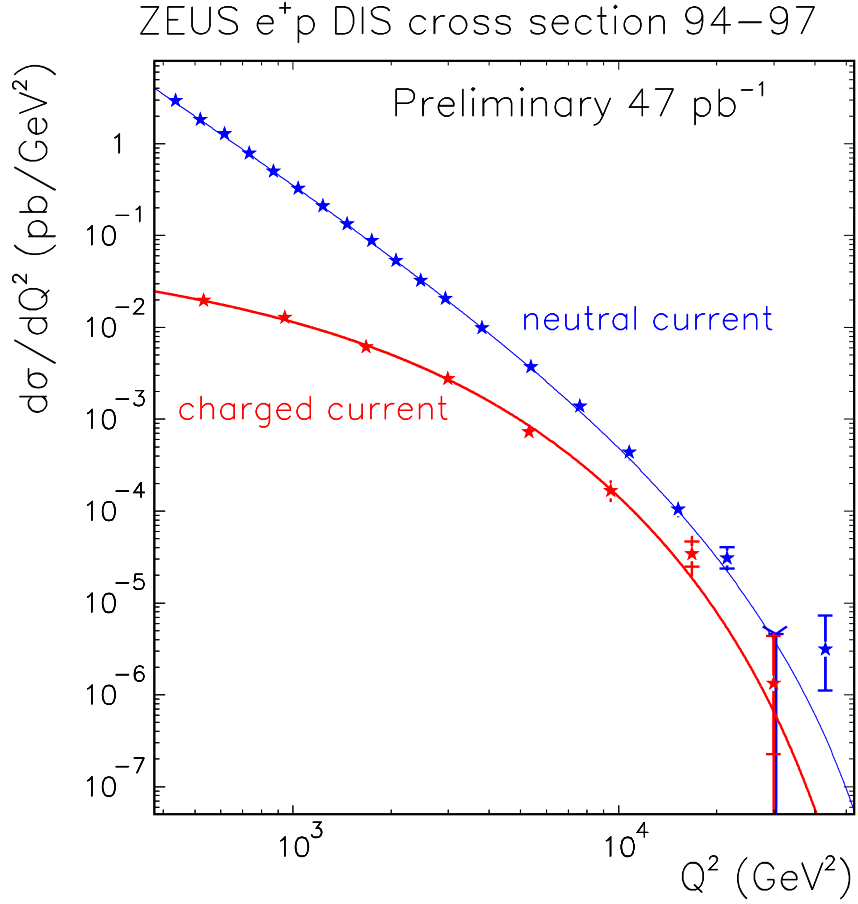,height=12cm}}
\caption{Charged and neutral current DIS cross sections at high
$Q^2$, as measured by the ZEUS collaboration \protect\cite{ZEUS}
in  $e^+p$ scattering at HERA.}
\end{center}
\label{fig:herahiq}
\end{figure}
In this talk we will concentrate mainly on the first of these, i.e. the impact
of CC and NC measurements on parton distributions and $\as$. Our
approach will be to examine the accuracy with which predictions, based
on global fits to DIS data at lower $Q^2$ and NLO QCD evolution to
higher $Q^2$, can already be made for the kinematic regime covered by HERA.
These predictions can then be used as a benchmark to assess the impact
of present and future HERA data. Since the issues are slightly different
for CC and NC cross sections, we will discuss each of these in turn.
Before doing so, for reference we collect together the leading-order
expressions for the relevant scattering cross sections:\footnote{In
these expressions the proton mass is set to zero, and $Q^2 = x y s$.}
\begin{itemize} 
\item[$\bullet$]\ \underline{neutral current} 
\beqn 
{d^2\sigma_{NC}(e^\pm p) \over d x d Q^2} &=&  {2 \pi\alpha^2 \over x Q^4}\;   
\Big[   [1+ (1-y)^2] F_2(x,Q^2) -y^2  F_L(x,Q^2)  \nonumber \\   
&& \mp 2y\big(1-y\big) xF_3 (x,Q^2)  \Big]  
\label{eq:nc}
\eeqn
\beqn 
F_2(x, Q^2)     &= &  \sum_q [xq(x,Q^2)   +  
 x\bar{q}(x,Q^2) ] \; A_q(Q^2) \nonumber \\ 
x F_3(x, Q^2)   &= &  \sum_q [xq(x,Q^2)   -  
 x\bar{q}(x,Q^2) ] \; B_q(Q^2)  
\label{eq:F23def}
\eeqn 
\beqn 
A_q(Q^2) &= & \; e_q^2 - 2 e_q v_e v_q P_Z +  
(v_e^2 + a_e^2) (v_q^2 + a_q^2) P_Z^2 \nonumber \\   
B_q(Q^2) &= & \; \qquad - 2 e_q a_e a_q P_Z + 4 v_e a_e  
v_q a_q  P_Z^2 \nonumber \\  
P_Z &=  & \; {Q^2 \over Q^2 + M_Z^2}\; {\sqrt{2} G_\mu M_Z^2\over 4 \pi \alpha} 
\label{eq:ABdef}
\eeqn 
\item[$\bullet$]\ \underline{charged current} 
\beqn 
{d^2\sigma_{CC}(e^-p) \over d x d Q^2}  & = &  [1-{\cal P}_e ]{ G_\mu^2 \over 
2\pi} \Big({M_W^2\over Q^2 + M_W^2}\Big)^2  \nonumber \\ 
& \times & \sum_{i,j}\;
\Big[  \vert V_{u_id_j}\vert^2  u_i(x,Q^2)   
  + (1-y)^2  \vert V_{u_j d_i}\vert^2 \bar d_i(x,Q^2)   \Big]  \nonumber \\ 
\label{eq:nce}
\eeqn
\beqn 
{d^2\sigma_{CC}(e^+p) \over d x d Q^2} &  = &  [1+{\cal P}_e ]{ G_\mu^2 \over 
2\pi} \Big({M_W^2\over Q^2 + M_W^2}\Big)^2 \nonumber \\ 
& \times & \sum_{i,j}\;
\Big[  \vert V_{u_id_j}\vert^2  \bar u_i(x,Q^2)   
  + (1-y)^2  \vert V_{u_j d_i}\vert^2   d_i(x,Q^2)   \Big]  \nonumber \\ 
\label{eq:ncp}
\eeqn 
\end{itemize} 
{}From these expressions we see that (i) the charged current
cross section is relatively suppressed by
$\cO(Q^4)$ at small $Q^2$ where the neutral
current cross section is dominated by photon exchange, and (ii) at very 
high $Q^2 \gg \cO(M_V^2)$, the charged and neutral cross sections are
of the same order. The HERA data confirm this 
behaviour: Fig.~\ref{fig:herahiq}
 shows the neutral and charged current
 cross sections, integrated over $x$, for  $e^+p$ scattering at high $Q^2$ 
 measured by ZEUS (see \cite{Andy}),
together with the Standard Model predictions.

\section{Neutral current cross sections}

\begin{figure}[tb]
\vspace{-0.75cm}                                                                
\begin{center}     
\epsfig{figure=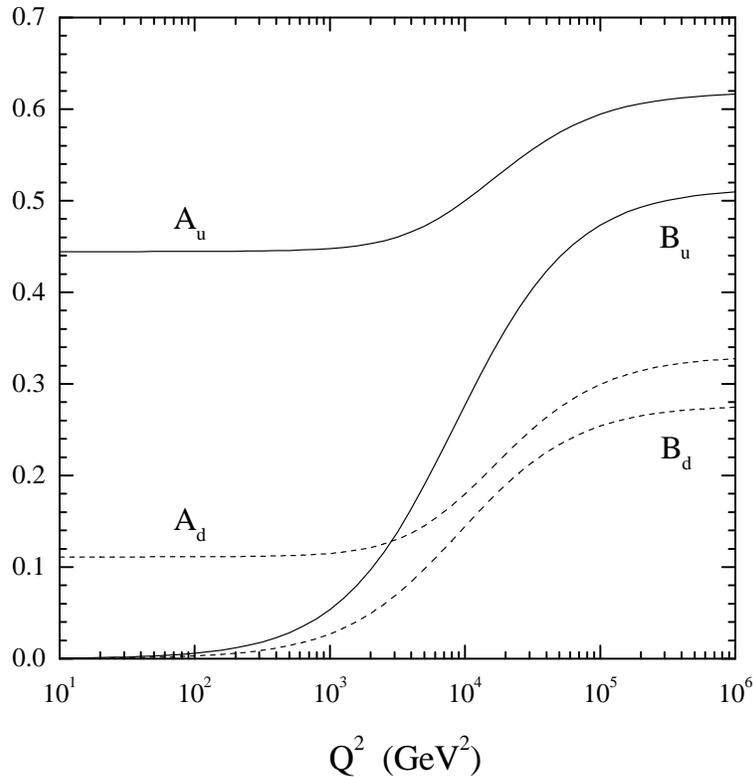,height=12cm}
\end{center}    
\vspace{-0.5cm}
\caption{The $Q^2$ dependence of the
parton combination functions $A_q$ and $B_q$ which appear
in the neutral current cross-section expressions of
Eq.~(\protect\ref{eq:F23def}).}
\label{fig:ABplot}
\end{figure}             
In QCD, the longitudinal structure function $F_L$ is suppressed
by $\cO(\asQ)$ compared to $F_2$ and $F_3$, and so at high $Q^2$
its contribution is numerically small. Ignoring overall factors, we
therefore have
\beqn
\sigma_{NC}(e^- p) + \sigma_{NC}(e^+ p) &\ \sim\ & \ F_2 =
x \sum_q A_q (q + \bar q)\; , \nonumber \\
\sigma_{NC}(e^- p) - \sigma_{NC}(e^+ p) &\ \sim\  & x F_3 =
x \sum_q B_q (q - \bar q) \; .\nonumber
\eeqn
The $Q^2$ dependence of these cross section
combinations (disregarding the overall $1/Q^4$) comes from two
sources: $Z^0$ propagator form-factor effects,
as contained in the $A_q$ and $B_q$,
and  logarithmic DGLAP \cite{DGLAP} evolution of the parton
distributions. Both are visible in current HERA data  \cite{Andy}.
Note that as $Q^2 \to 0$, $A_q \to e_q^2$ and $B_q \to 0$. Thus
$F_2$ is the {\it same} structure function as measured in fixed-target
experiments at lower $Q^2$. The $Q^2$ dependence of the $A_q$ and $B_q$ for
$u$-- and $d$--type quarks is illustrated in Fig.~\ref{fig:ABplot}.
The point to note here is that the relative mix of the two quark types
does not change radically as $Q^2$ increases --- up quarks still dominate
at high $Q^2$. This implies that the uncertainty in the extrapolation
of, say, $F_2^{\mu p}$ from low to high $Q^2$ at large $x$ from changes
in the relative contributions of the valence $u$ and $d$ quarks is very small.

\begin{figure}[tb]
%\vspace{-0.5cm}                                                                
\begin{center}     
\epsfig{figure=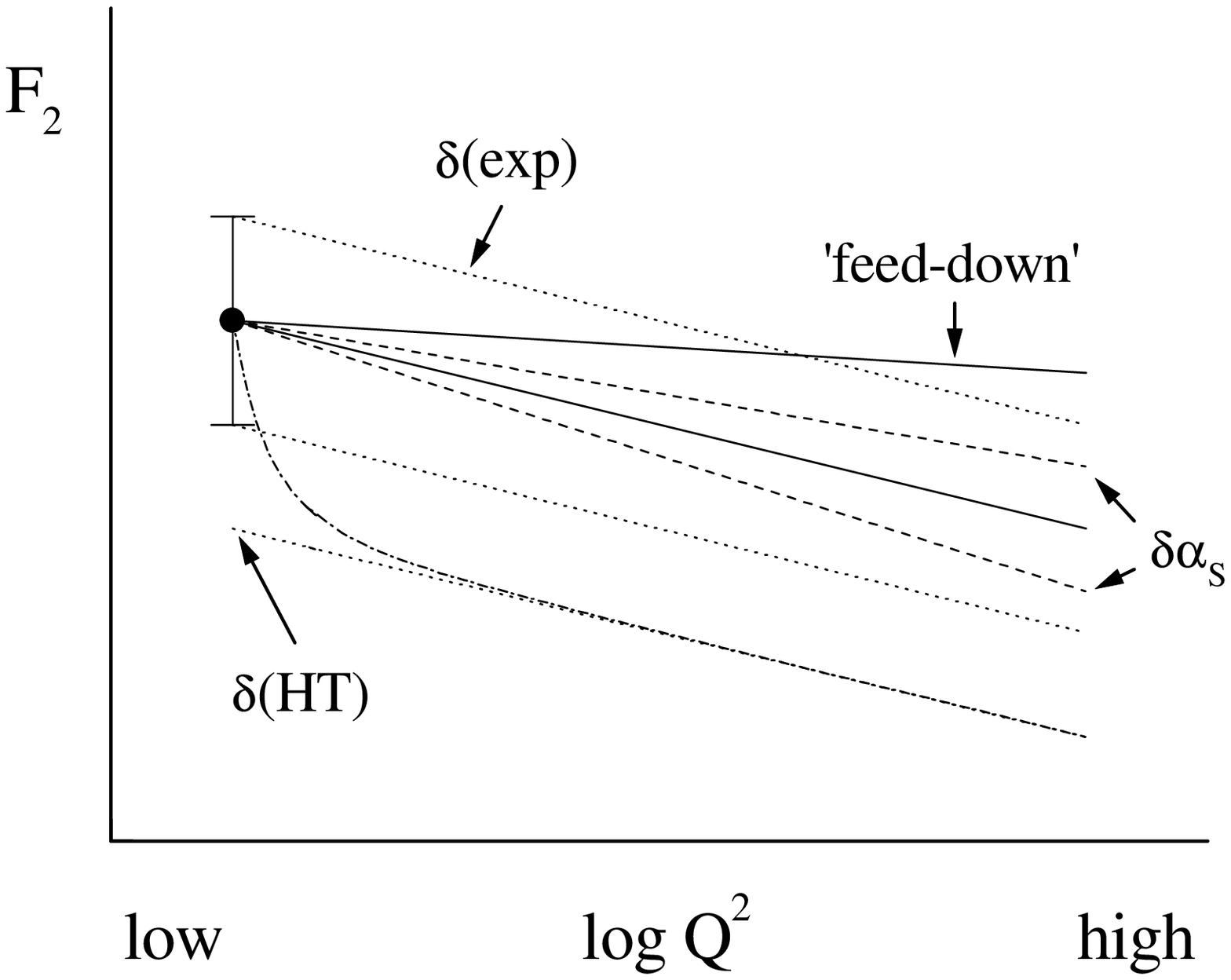,height=9cm}
\end{center}
\vspace{-0.5cm}    
\caption{Illustration of the
different contributions to the uncertainty in the prediction of
$F_2$ at high $Q^2$, at a fixed (large) value of $x$,
given a measurement at lower $Q^2$.}
\label{fig:ecartoon}
\end{figure}
\begin{figure}[tb]
\vspace{-1.0cm}                                                                
\begin{center}     
\epsfig{figure=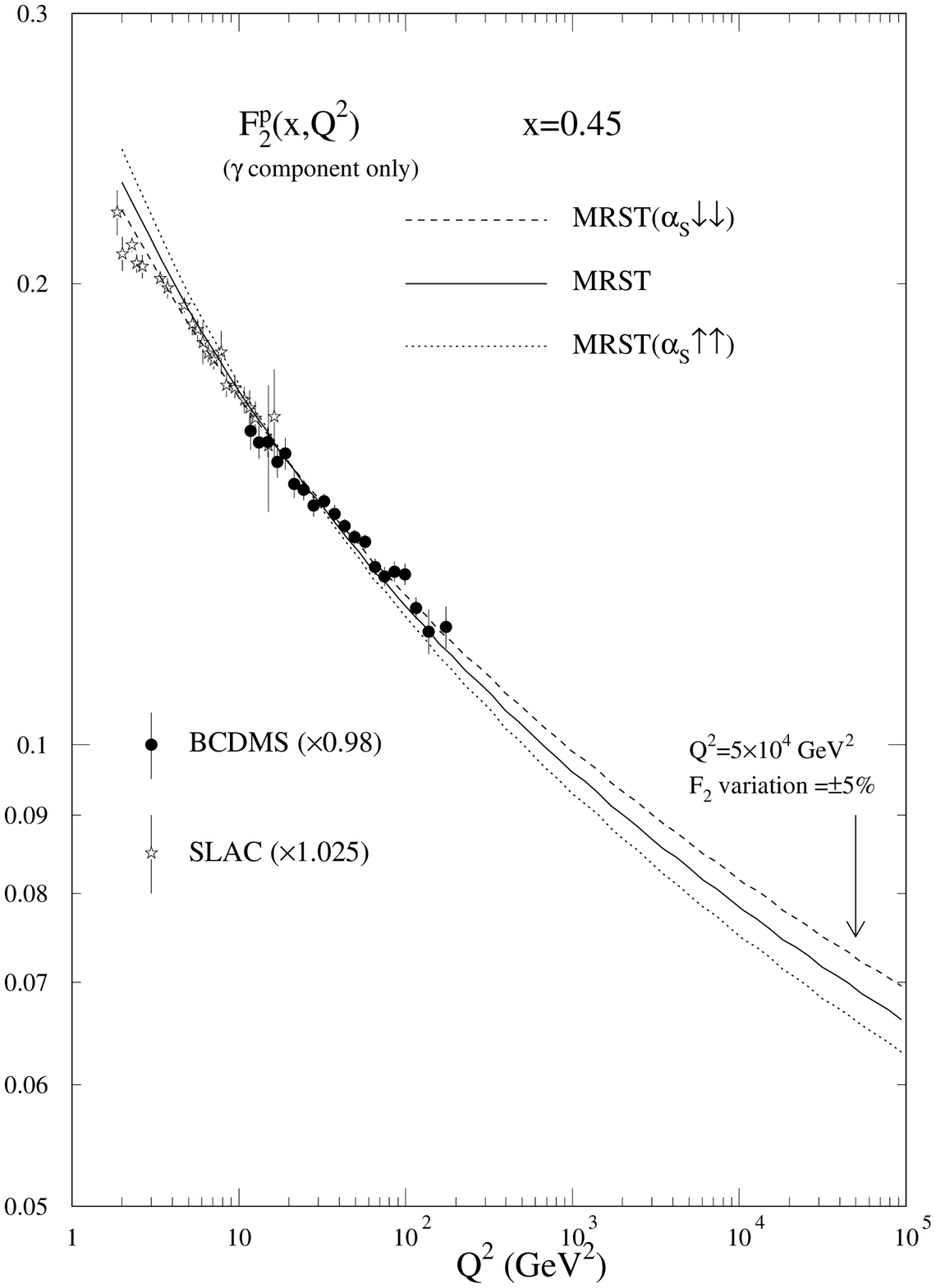,height=14cm}    
\end{center}
\vspace{-0.5cm}    
\caption{The extrapolation of the fits at $x=0.45$ to high $Q^2$ using  the    
MRST, MRST($\asup$) and MRST($\asdown$) sets of partons, from 
Ref.~\protect\cite{MRST}.}
\label{fig:x45plot}
\end{figure}

             
Given a measurement of $F_2$ at lower $Q^2$, how well can we then
predict $F_2$ in the high--$Q^2$ region probed by HERA?
Figure~\ref{fig:ecartoon} is a schematic (i.e. not-to-scale)
illustration of the largest sources of uncertainty.\footnote{We are
assuming here that the electroweak parameters associated with the $Z^0$
exchange contribution are already very precisely known from LEP
measurements.}  First, any
measurement error on the low--$Q^2$ data propagates directly through to
high $Q^2$. For fixed-target DIS data at medium-large $x$, this
uncertainty is of order $\pm 3\%$ (see for example \cite{MRST}).
Second, any uncertainty on $\asQ$ affects the evolution of $F_2$ via the
large-$x$  DGLAP equation
\beq
{\partial F_2 \over \partial \log Q^2} \sim \asQ\; P^{qq} \otimes F_2 \;
.
\label{eq:DGLAP}
\eeq
The effect on the evolution of a `world average' value and error,
$\asmz = 0.1175 \pm 0.005$, is illustrated in
Fig.~\ref{fig:x45plot}, taken from
Ref.~\cite{MRST}. Evidently the error on $\alpha_s$  induces an uncertainty of
order $\pm 5\%$ in $F_2$ at high $Q^2 \sim 10^5\; \GeV^2$.

An error in the evolution of $F_2$ could also be made if there is a
significant higher-twist contribution to the low--$Q^2$ data that is
not taken account in the fitting and subsequent evolution. This is
potentially a problem at very large $x$, since the higher-twist
contributions are expected to behave as $1/(1-x)Q^2$ relative to
the leading-twist contribution. It is difficult to pin down the precise
size of this effect --- most  analyses apply  a minimum cut in $W^2 =
(1-x)Q^2/x$ to fixed-target data and fit the remaining data using
leading-twist NLO DGLAP. A recent study to quantify the impact of a
possible higher-twist contribution on extracted parton distributions is
reported in \cite{MRSTHT}.

Finally, the structure function $F_2$ in the convolution on the
right-hand side of the DGLAP equation  (\ref{eq:DGLAP}) is sampled at
$x' \geq x$. The evolution is therefore susceptible to the `feed-down' of
an anomalously large contribution to $F_2$ at $x \approx 1$. Such a
contribution could escape detection by the fixed-target measurements
while still influencing the evolution of $F_2$ to the HERA region, see
for example the study of Ref.~\cite{TUNG}.  Again, it is hard to
quantify the maximum effect that such an anomaly could have on $F_2$ at
high $Q^2$. Certainly in global fits  that adopt the
physically reasonable assumption that (leading-twist) $F_2$ decreases
smoothly to zero as $(1-x)^n$, with $n \simeq 3 - 4$ at low $Q^2$,
there is no uncertainty in the evolution of $F_2$ from the large-$x$
`unmeasured' region.

In summary, if higher-twist contributions have been correctly estimated
and if there  is no anomalous contribution to $F_2$ at very high x,
then we should be able to predict the high--$Q^2$ neutral current cross
sections at HERA to within about $\pm 5\%$, with the main uncertainty
appearing to come from the error on $\as$. A HERA measurement at this
level of precision would therefore provide a powerful check of the
theoretical technology based on leading-twist NLO DGLAP evolution.
If there is agreement between the low-- and high--$Q^2$ data sets, 
the latter can be incorporated into global fits to help pin down further
the parton distributions and $\as$.    Conversely, any gross 
deviation from the theoretical predictions could signal new physics.

\section{Charged current cross sections}

\begin{figure}[tb]
%\vspace{-0.5cm}                                                                
\begin{center}     
\epsfig{figure=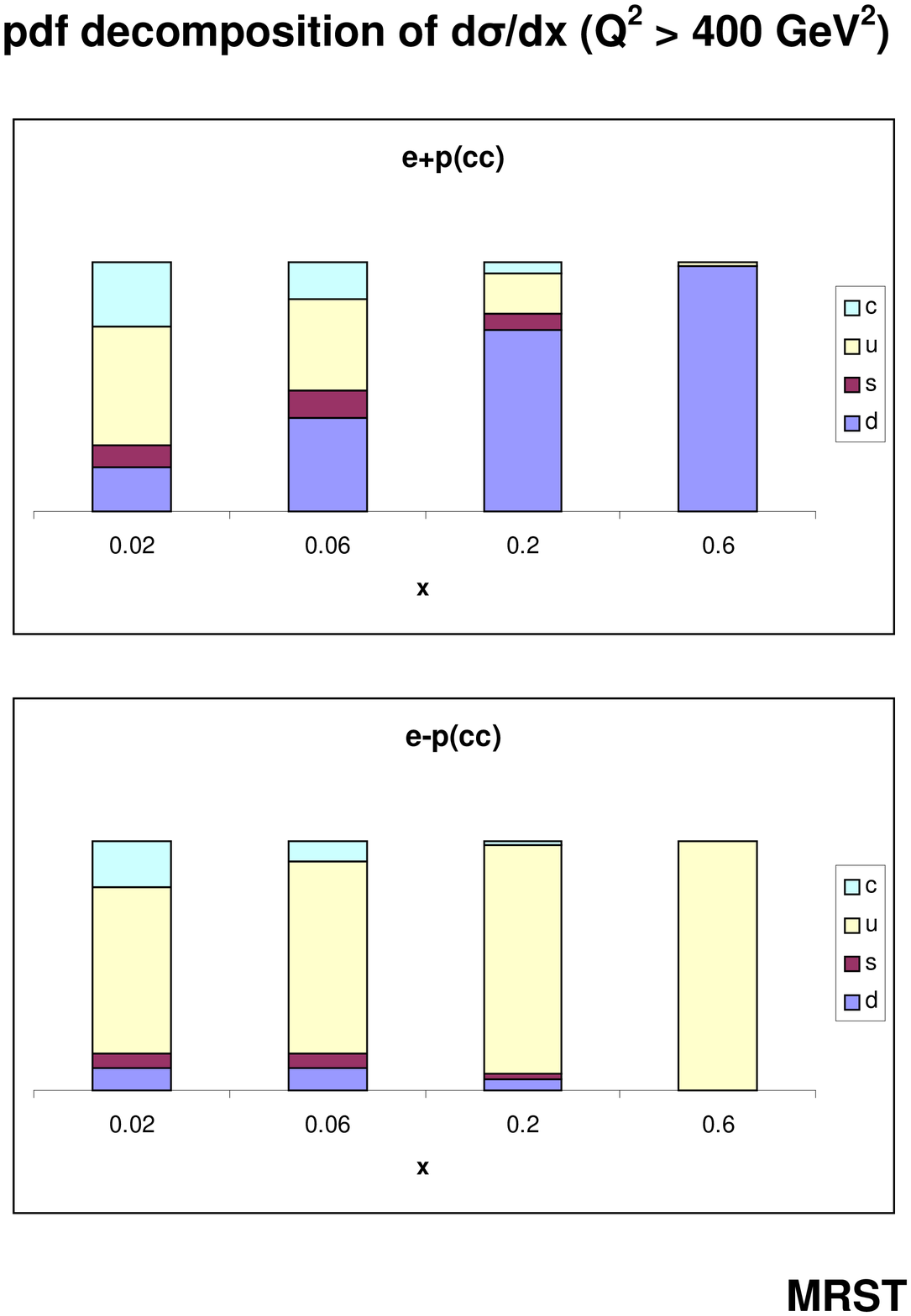,height=14cm}
\end{center}    
\vspace{-0.5cm}
\caption{Parton decomposition of the high--$Q^2$ $e^-p$ and $e^+p$ CC
cross sections.}
\label{fig:cchart}
\end{figure}
\begin{figure}[htb]
%\vspace{-0.5cm}                                                                
\begin{center}     
\epsfig{figure=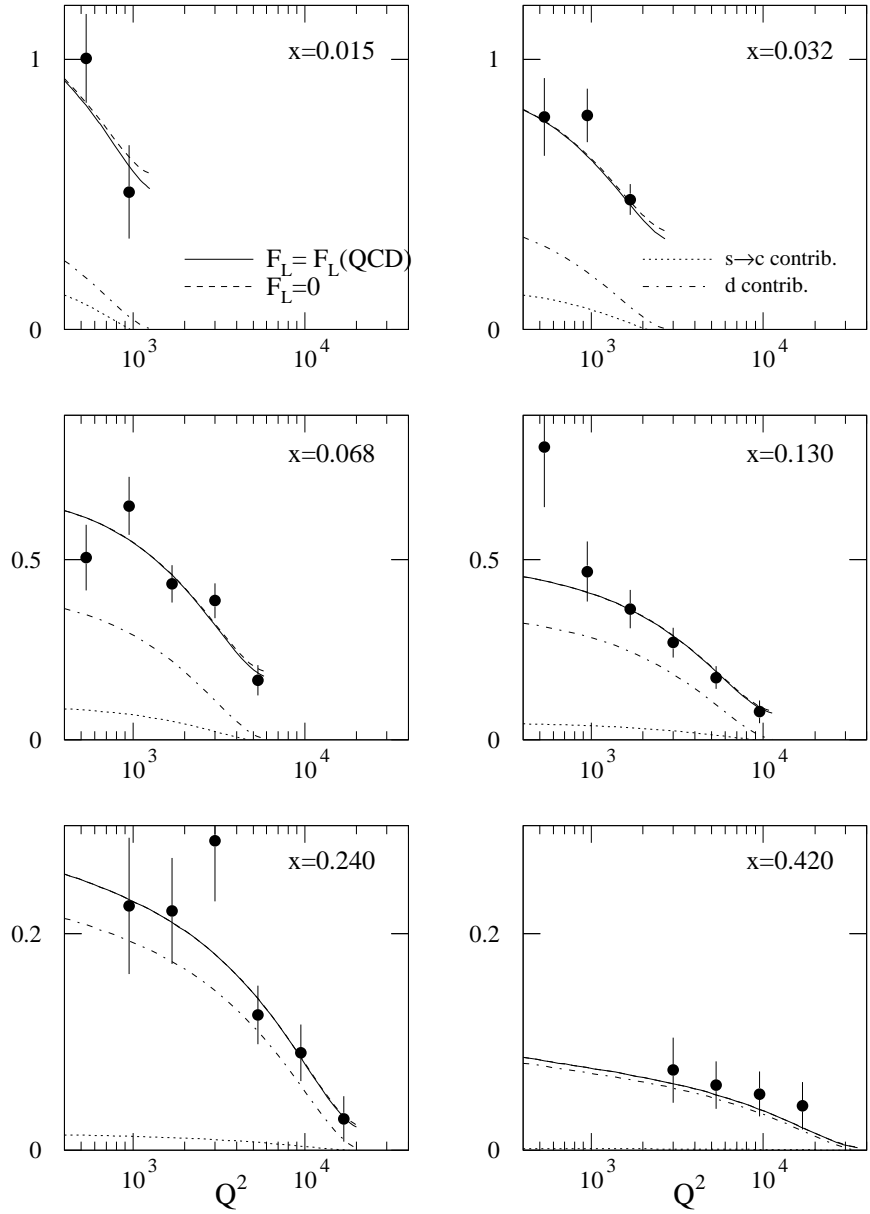,height=18cm}    
\end{center}
\vspace{-0.5cm}    
\caption{Comparison of the predictions \protect\cite{RGR}
 for  charged current
$e^+p$ cross sections using MRST partons \protect\cite{MRST}, with data from the ZEUS collaboration  \protect\cite{ZEUS}.}
\label{fig:zeusplot}
\end{figure}

The normalisation and $Q^2$ dependence of the charged current cross
sections (\ref{eq:nce},\ref{eq:ncp})
are, in principle, sensitive to the electroweak parameters
$G_\mu$ and $M_W$. The current and projected precision on the extraction
of these parameters was discussed at some length in the Working Group,
see \cite{WGi}. Notice, however, that there is also potentially useful
information on  parton distributions, since the flavour
decomposition is quite different from that of the neutral current cross
sections. Ignoring overall couplings, we have
\beqn
\sigma_{CC}(e^+ p)   &\ \sim\ & \ \ubar + \cbar + (1-y)^2 (d+s)
\longrightarrow (1-y)^2 d \; ,\nonumber \\
\sigma_{CC}(e^- p)   &\ \sim\ & \ u + c + (1-y)^2 (\dbar+\sbar) 
\longrightarrow u \; ,\nonumber
\eeqn
where the $x \to 1 $ limit is indicated. The quantitative breakdown
is illustrated in
Fig.~\ref{fig:cchart}, which
 shows the pdf decomposition of the $e^+p$ and $e^-p$  CC cross
sections as a function of $x$ at high $Q^2$.
Evidently the $e^-p$  cross section is completely dominated by the 
$u$--quark distribution and, as such, should be predictable with high
precision, assuming of course the
validity of DGLAP evolution as discussed in the previous section.
More interesting is the $e^+p$  cross section. This is 
dominated by the $d$--quark distribution at large $x$ (though not to the
same extent as the $u$ distribution dominates the $e^-p$ cross
section).  In Fig.~\ref{fig:cchart}, 74\% and 98\% of the leading-order
cross section comes from $e^+d$ scattering at $x=0.2$ and $0.6$
respectively.
The ratio $\sigma_{CC}(e^+p)/\sigma_{CC}(e^-p)$ therefore provides a
good measure of the $d/u$ ratio. Current information on
$d/u$ at large $x$ comes from fixed target $F_2^{\mu n}/F_2^{\mu p}$
measurements and the lepton asymmetry 
in $p \bar p \to W^\pm + X$, see for example \cite{MRST}. 
In the MRST fit,
NMC $n/p$ data are used to constrain the large-$x$ $d$--quark pdf in this way.
The corresponding predictions for $\sigma_{CC}(e^+p)$ are compared with
the ZEUS data \cite{ZEUS} in Fig.~\ref{fig:zeusplot} \cite{RGR}.
Although the agreement is entirely satisfactory, there is  some
evidence of a slight excess of data over theory in the largest $x$ 
($=0.42$) bin.
Could this imply that the $d/u$ ratio is being underestimated in the
standard global fits? Any attempt to increase $d/u$ at large $x$ 
in the global fit leads
to a direct conflict with the $n/p$ data. However, Bodek and Yang have
argued \cite{BODEK}
that the latter should be corrected for nuclear binding effects which,
at large $x$, lead to a larger $d/u$ ratio, in `better' agreement
with the ZEUS data. This is an issue that deserves more attention, and
improved precision on the HERA $e^+p$ data would be very valuable.

\section{Summary}

In this brief review we have highlighted some of the physics issues
relating to  neutral and charged current cross sections
at high $x$ and $Q^2$ at HERA. Although there is some scope for
obtaining information on electroweak parameters, in 
particular $M_W$ \cite{WGi},
the main impact of future data is likely to be in testing perturbative
QCD evolution via the DGLAP equation and in obtaining information on the pdfs.
The $d$--quark distribution, for example, is directly probed by the
charged current $e^+p$ cross section. 
Finally, we note that the HERA
high $x,Q^2$  DIS kinematic region overlaps with the corresponding
region that
will be probed by many hard scattering processes at the LHC.

\end{document}